\documentclass[dvips]{article}
\usepackage{icrctc07}

\title{A prototype device for acoustic neutrino detection in Lake Baikal}
\shorttitle{Underwater 4-channel digital device}
\authors{K. Antipin$^{1}$, V. Aynutdinov$^{1}$, V. Balkanov$^{1}$, 
I. Belolaptikov$^{4}$, D.Bogorodsky$^{2}$, N.Budnev$^{2}$, I. Danilchenko$^{1}$, G. Domogatsky$^{1}$,
A. Doroshenko$^{1}$, A. Dyachok$^{2}$,  Zh.Dzhilkibaev$^{1}$, S. Fialkovsky$^{6}$,
O. Gaponenko$^{1}$, K. Golubkov$^{4}$, O. Gress$^{2}$, T. Gress$^{2}$, O. Grishin$^{2}$, A. Klabukov$^{1}$,
A. Klimov$^{8}$, A. Kochanov$^{2}$, K. Konischev$^{4}$, A. Koshechkin$^{1}$, V. Kulepov$^{6}$, L. Kuzmichev$^{3}$,
E. Middell$^{5}$, S. Mikheyev$^{1}$, M. Milenin$^{6}$, R. Mirgazov$^{2}$, E. Osipova$^{3}$, G. Pan'kov$^{2}$,
L. Pan'kov$^{2}$, A. Panfilov$^{1}$, D. Petukhov$^{1}$, E. Pliskovsky$^{4}$, P. Pokhil$^{1}$, V. Poleshuk$^{1}$,
E. Popova$^{3}$, V. Prosin$^{3}$, M. Rosanov$^{7}$, V. Rubtzov$^{2}$, B. Shoibonov$^{4}$, A. Sheifler$^{1}$,
A. Shirokov$^{3}$, Ch. Spiering$^{5}$, B. Tarashansky$^{2}$, R. Wischnewski$^{5}$, I. Yashin$^{3}$,
V. Zhukov$^{1}$
}
\shortauthors{V. Aynutdinov et al}
\afiliations{$^1$Institute for Nuclear Research, Moscow, Russia\\
$^2$Irkutsk State University, Irkutsk, Russia\\
$^3$Skobeltsyn Institute of Nuclear Physics  MSU, Moscow, Russia\\
$^4$Joint Institute for Nuclear Research, Dubna, Russia\\
$^5$DESY, Zeuthen, Germany\\
$^6$Nizhni Novgorod State Technical University, Nizhni Novgorod, Russia\\
$^7$St Petersburg State Marine University, St Petersburg, Russia\\
$^8$Kurchatov Institute, Moscow, Russia
}
\email{nbudnev@api.isu.ru}
\abstract{In April $2006$ a $4$-channel acoustic antenna has been put in long-term operation on Lake Baikal. The detector was installed  at a depth of about $100$ m on the instrumentation string of Baikal Neutrino Telescope NT200+. This detector may be regarded as a prototype of subunit for a future underwater acoustic neutrino telescope. We describe the design of acoustic detector and present first results obtained from data analysis.}

\begin{document}
\maketitle

\section{Introduction}
\begin{figure*}[t]
\begin{center}
\includegraphics [width=0.98\textwidth]{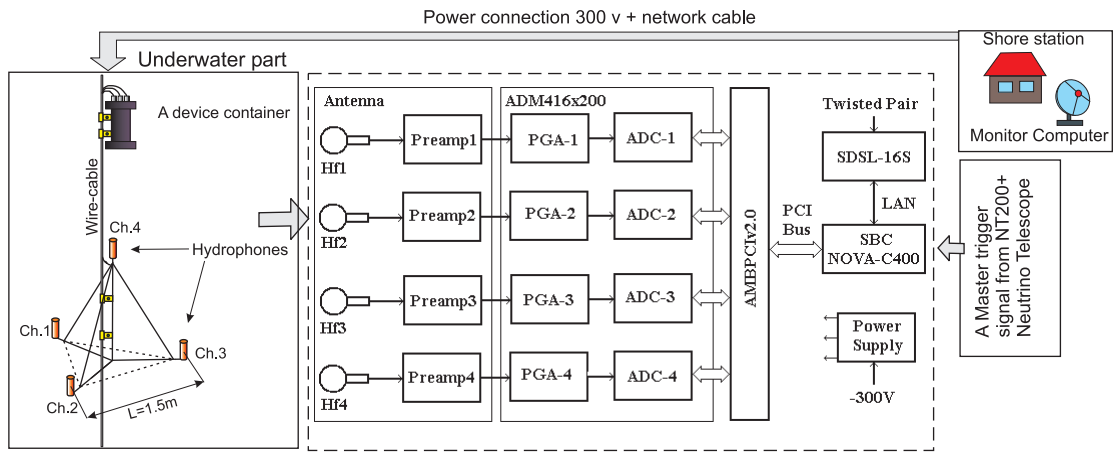}
\end{center}
\caption{Schematic view of  underwater 4-channel digital device for detection of acoustic signals from high energy neutrinos.}
\label{fig1}
\end{figure*}

The large scale neutrino telescopes currently under operation (NT200+ in Lake Baikal, AMANDA/IceCube at the South Pole and ANTARES in the Mediterranean) detect the Cherenkov light emitted in water or ice by relativistic charged particles produced via neutrino interactions with matter. Back in 1957, G.A. Askaryan has shown that a high-energy particle cascade in water should also produce an acoustic signal \cite{Askarian-1957}. The absorption length for acoustic waves with a frequency about 30 kHz (the peak frequency of acoustic signals from a shower) in sea water is at least an order of magnitude larger than that of Cherenkov radiation, in the fresh Baikal water this ratio is even close to 100 \cite{Clay}. Therefore acoustic pulses can be detected from considerably larger distances than Cherenkov radiation, and the acoustic method appears to be attractive for the detection of ultra high-energy neutrinos \cite{Askarian-1977}. However, the technology of acoustic detection in high-energy physics is much worse developed than optical methods. Since several years, however, an increasing number of feasibility studies on acoustic particle detection are performed \cite{ARENA}.  

In order to test the possibility of acoustic detection of high-energy neutrinos in 
Lake Baikal, the Baikal collaboration started with an in-situ study of acoustic noise 
which constitutes the background for the acoustic neutrino detection in the lake. For the purpose of noise measurement, an autonomous hydro-acoustic recorder with two input channels has been developed. We have performed a series of hydro-acoustic measurements in Lake Baikal in order to investigate the the background properties  \cite{Zeuthen-1, Akustika-noise}. It turned out that at stationary and homogeneous meteorological conditions the integral noise power in the frequency range $20$-$50$ kHz can reach levels as low as about $1$ mPa. At the same time, short acoustic pulses with different amplitudes and shapes including bipolar ones have been 
observed. The latter should be considered as a background for acoustic neutrino detection. However,  the overwhelming majority of the short pulses have probably been generated by quasi-local sources or are due to interference of noise sound waves coming from a layer near the surface.

Taking into account these properties of the noise, we conclude that the most promising way to detect acoustic particle signals is to deploy a net of rather compact acoustic antennas at relatively shallow depths (for example about $100$--$200$ m for Lake Baikal) and monitoring the water volume top-down. It is also necessary to suppress signals from the surface by caps made of a sound-absorbing material and mounted on top of the antennas.

\section{Device for the detection of acoustic signals from high energy neutrinos}

Cascades generated by neutrino interactions in water should produce bipolar acoustic impulses with $30$--$50$ $\mu$sec duration. Most of the acoustic signal energy is concentrated within a disk which's axis coincides with the cascade direction \cite{Askarian-1979, Learned-1979, Dedenko}. Disk shape and bipolarity are therefore the basic signatures to search for. Extraction of small signals from background requires an antenna consisting of a set of hydrophones. The optimum distance between the hydrophones is defined by the condition that it must safely exceed several 
wave-lengths of the expected signal but, on the other hand, should not be too large in order to minimise the number of background impulses captured within the coincidence time window. We have constructed a digital hydro-acoustic device with four input channels shown in Fig. \ref{fig1} \cite{Zeuthen-2}. 

The module was designed for common operation with the Baikal Neutrino Telescope NT200+ and has been installed in April 2006 at one of the moorings of NT200+. To suppress the amount of raw information to be transferred to the shore station, data are pre-processed at the deep site using an algorithm described in \cite{ICRC29}.

There are three regimes of operation of the instrument:
\begin{enumerate}
\item{ Transmitting of a one-second sample of data from all hydrophones to the shore computer centre, after a trigger signal from  NT200+.} 
\item{Online search for short acoustic pulses, 
which can be interpreted as signals from distant quasi-local sources.} 
\item{An autonomous analysis of acoustic background statistics.}
\end{enumerate}
The joined operation with NT200+ might give us an opportunity to identify the properties of acoustic emission from cascades and provide an energy calibration (assuming that signal strength and flux are high enough and the energy threshold low enough to collect a usable number of true coincidences). 

\begin{figure}[t]
\begin{center}
\includegraphics [width=0.48\textwidth, trim= 0.9cm 1.2cm 0cm 0.4cm]{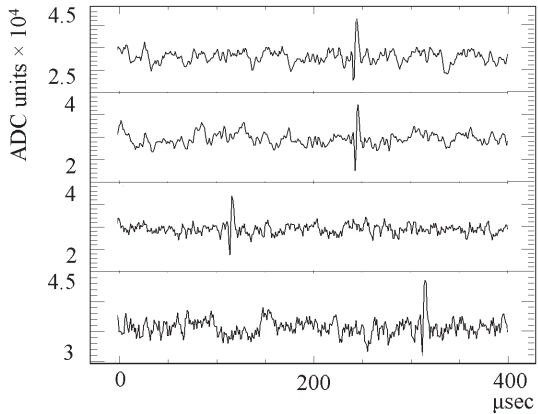}
\caption{Sample of detected bipolar pulse.}\label{fig2}
\end{center}
\end{figure}

\section{Results}

The off-line analysis includes the following steps:

\begin{enumerate}
\item{
All data have been arranged in time frames, one per hydrophone,
in a way that all data which may belong to a common
candidate event are covered by all four time frames.
The frames start at the same time and have a 
minimum length of 2 $\tau_{cross}$, where 
$\tau_{cross}$ is the maximum 
duration for a plane wave front to cross the antenna
volume.} 
\item{Search for segments, in which the 
signal amplitude exceeds the dispersion $\sigma$
of the acoustic noise in the corresponding time frame
by a factor $k$. For this analysis, $k$=2.5.}
\item{Neighboured segments with amplitude larger
than $k\times \sigma$ and opposite sign are
combined, with the additional condition that 
the interval between these segments 
is  shorter than the added length of both
segments.}
\item{Pulse form classification according to the
number of combined segments into bipolar, 3-polar 
etc. pulses.}
\item{Pulse duration cut, $\tau < 100 \mu$sec for this analysis.}
\item{Test of pulse forms to be described 
by a function $F(x)$ that approximates 
the  bipolar pulses expected from a high-energy shower. 
Data approximation criterion: 
$ \sum_{i=1}^{m}(F(t_i) - A(t_i))^2 < N$, $N=1.5$ in this paper.}

\end{enumerate}

\begin{figure}[t]
\begin{center}
\includegraphics [width=0.49\textwidth, trim= 0.4cm 0.5cm 0.5cm 1cm]{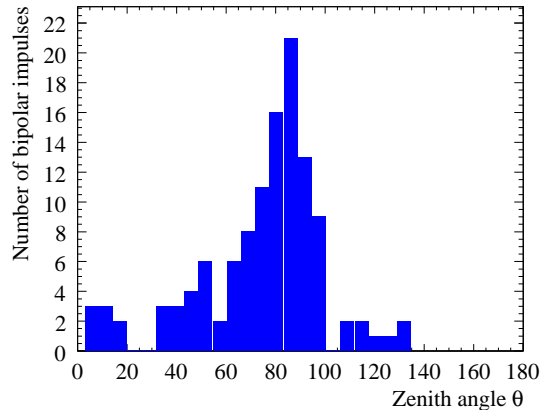}
\caption{The distribution of bipolar pulses versus zenith angle.}\label{fig3}
\end{center}
\end{figure}

Fig. \ref{fig2} presents an example of a bipolar pulse, 
selected by this procedure. Fig. \ref{fig3} shows the zenith angle distribution of 
bipolar pulses registered in April-May 2006. Most pulses are located in the 
vicinity of the horizontal plane. Sources of pulses coming from just below
horizon are likely located also in the near-surface zone. They appear to come
from below horizon due to refraction which is caused by the growth in sound velocity with depth. From the region  $\pm 45^o$ around the opposite zenith, no
event with bipolar pulse form has been observed.

\begin{figure}
\begin{center}
\includegraphics [width=0.5\textwidth, trim= 0.4cm 0.5cm 0.5cm 1cm]{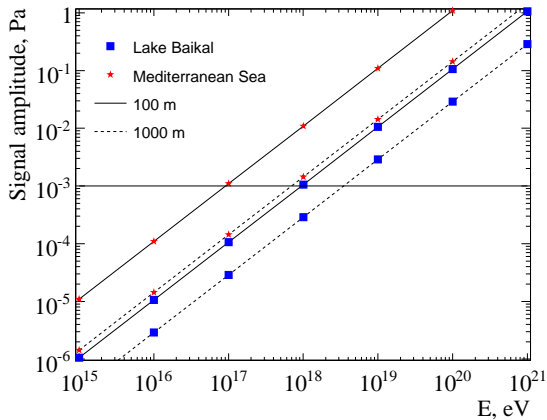}
\end{center}
\caption{Dependence of expected acoustic signals amplitude generated by high-energy showers versus shower energy.}
\label{fig4}
\end{figure}

Fig. \ref{fig4} presents a comparison between amplitudes calculated using
the formula from \cite{Askarian-1979, Chen, Weiss}
for distances $100$  and $1000$ m in Mediterranean Sea and in Lake Baikal.
The high-frequency noise level in Lake Baikal can be
as low as $\sim 1$ mPa (horizontal line in Fig. \ref{fig4}),
and is comparable
to ~3 mPa ambient acoustic noise for calm sea \cite{niess}.
It is possible to register showers of energy higher
than $10^{18}$ eV from a distance of $100$ m,
and higher than $10^{19}$ eV from $1000$ m. At distances
much larger than $1$ km, signal amplitudes of showers in
Baikal water are similar or higher than in the Mediterranean
due to lower sound absorption in freshwater.

\section{Summary and Outlook}
The results of the experiment have demonstrated the feasibility of the proposed acoustic pulse detection technique in searching signals from cascade showers. Although the Baikal water temperature is close to the temperature of its maximal density, the absence of strong acoustic noise sources in the lake's deep zone, and the very low absorption 
of sound in freshwater may result in neutrino detection in Lake Baikal with a threshold as low as $10^{18}$ -- $10^{19}$ eV. 

This motivates further activities towards a large-scale acoustic neutrino 
detector in Lake Baikal.

\section{Acknowledgements}
This work is partially supported by Russian Ministry of Education and Science, Russian Fund of Basic Research (grants 07-02-00791, 07-02-10009, 07-02-10013, 05-02-16593,  05-02-17476) and by the grants of Irkutsk State University (111-02-000/7-06, 111-02-000/7-07, 111-02-000/7-09).

\end{document}